\documentclass[aps,twocolumn,prl,superscriptaddress]{revtex4-2}
\usepackage{amsmath}
\usepackage{amssymb}
\usepackage{amsfonts}
\usepackage{color}
\usepackage{graphicx}
\usepackage{bm}
\usepackage[export]{adjustbox}

\newcommand {\Wi}{W\!i}

\newcommand{\revision}{\color{black}}
\newcommand{\revisiontwo}{\color{black}}

\begin{document}

\title{\revisiontwo{Coherent structures in plane channel flow of dilute polymer solutions with vanishing inertia}}

\author{Alexander Morozov}
\email{alexander.morozov@ed.ac.uk}
\affiliation{SUPA, School of Physics and Astronomy, The University of Edinburgh, James Clerk Maxwell Building, Peter Guthrie Tait Road, Edinburgh, EH9 3FD, United Kingdom}

\begin{abstract}
When subjected to sufficiently strong velocity gradients, solutions of long, flexible polymers exhibit flow instabilities and chaotic motion, often referred to as elastic turbulence. Its mechanism differs from the familiar, inertia-driven turbulence in Newtonian fluids, and is poorly understood. Here, we demonstrate that the dynamics of purely elastic pressure-driven channel flows of dilute polymer solutions are organised by exact coherent structures that take the form of two-dimensional travelling waves. Our results demonstrate that no linear instability is required to sustain such travelling wave solutions, and that their origin is purely elastic in nature. We show that the associated stress profiles are characterised by thin, filament-like arrangements of polymer stretch, which is sustained by a solitary pair of vortices. We discuss the implications of the travelling wave solutions for the transition to elastic turbulence in straight channels, and propose ways for their detection in experiments.
\end{abstract}

\maketitle

Close to the transition to chaos, a wide range of systems, spanning magnetoelastic devices \cite{Ditto1990}, electric circuits \cite{Hunt1991}, chemical reactions \cite{Petrov1993}, flows of Newtonian fluids \cite{Tuckerman2020,Linkmann2015}, neural and myocardial \cite{Garfinkel1992} tissues, and lasers \cite{Roy1992}, exhibits low-dimensional dynamics organised by unstable coherent structures. The phase space vicinity of such structures is generally attractive, with a (relatively) small number of unstable directions. In this scenario, a chaotic trajectory amounts to a `pin-ball'-like motion in phase space that spends a long time in close vicinity of coherent structures. Knowledge of such structures thus allows for relatively accurate statistical description of an otherwise unpredictable system \cite{Auerbach1987}.

In Newtonian parallel shear flows, the discovery of travelling waves and periodic orbits organised around streamwise streaks and vortices has lead to the prediction of the transitional values of the Reynolds number \cite{Waleffe1997}, revealed the transient nature of turbulence near its onset \cite{Hof2006,Avila2011}, and helped to identify directed percolation as the universality class for the transition to Newtonian turbulence \cite{Sano2016,Lemoult2016,barkley_2016}.
Addition of small amounts of high-molecular weight flexible polymers was shown to suppress Newtonian coherent structures in parallel shear flows \cite{Stone2002,Dubief2004}, while higher levels of viscoelasticity trigger elasto-inertial turbulence \cite{Samanta2013,Choueiri2018,Lopez2019,Yamani2021}, characterised by a different set of coherent structures \cite{Choueirie2021,Shekar2019,Shekar2020,Shekar2021}.

At very low Reynolds numbers, in the presence of sufficiently strong velocity gradients, dilute solutions of high-molecular weight flexible polymer molecules exhibit a unique chaotic flow state often referred to as elastic turbulence \cite{Groisman2000,Datta2021preprint}. Its origins differ from the inertia-dominated turbulence in Newtonian fluids, and lie in the flow-induced stretch and orientation of polymer molecules \cite{Steinberg2021}. Despite being implicated as a major production-limiting factor in polymer processing \cite{Denn2001}, elastic turbulence is still poorly understood.

Here, we shed light on the transition to elastic turbulence in parallel shear flows of polymer solutions, like flow through a plane channel or a straight pipe.  For the parameters relevant to typical experiments with dilute polymer solutions, parallel shear flows are linearly stable \cite{Datta2021preprint,Sanchez2022}. It has previously been proposed
theoretically \cite{Morozov2007,Morozov2019} that such flows exhibit a direct, sub-critical transition to elastic turbulence, similar to their Newtonian counterparts. These predictions are supported by experiments in straight microfluidic channels that report the existence of strong velocity fluctuations that appear sub-critically above the onset flow rate \cite{Bertola2003,Bonn2011,Pan2013,Qin2017,Qin2019}. The emerging transition scenario suggests strong parallels between purely elastic and Newtonian flows \cite{Morozov2007}, implicating coherent structures in organising the phase-space dynamics of both systems. Yet, very little is known about purely elastic flow structures experimentally \cite{Qin2019,jha2020preprint}, and we are not aware of any numerical simulations of such flows to date. Despite the aforementioned advances, purely elastic turbulence is still being commonly rationalised in terms of the dynamics of individual polymer molecules in imposed random flows \cite{Steinberg2021}. In this Letter, we demonstrate numerically the existence of two-dimensional travelling wave solutions (TWS) in parallel shear flows of dilute polymer solutions at vanishingly small Reynolds numbers and discuss how they are related to turbulent dynamics in such flows.

We perform direct numerical simulations of two-dimensional pressure-driven channel flow of a model simplified Phan-Thien-Tanner (PTT) polymeric fluid \cite{PhanThien1977}, chosen to capture the shear-thinning nature of dilute polymer solutions \cite{Datta2021preprint}. We performed linear stability analysis and confirmed that the flow is linearly stable for all parameters studied here (see \cite{SI} for detail). We consider a straight channel formed by two infinite parallel plates, with $x$ and $y$ being Cartesian coordinates along the streamwise and gradient directions, respectively. Equations are rendered dimensionless by using $d$, $\mathcal{U}$, $d/\mathcal{U}$, and $\eta_p \mathcal{U}/d$, and $(\eta_s+\eta_p)\mathcal{U}/d$ as the units of length, velocity, time, stress, and pressure, respectively. Here, $d$ is the channel half width, $\eta_s$ and $\eta_p$ are the solvent and polymeric contributions to the viscosity, and $\mathcal{U}$ is the maximum value of the laminar fluid velocity of a Newtonian fluid with the viscosity $\eta_s+\eta_p$ at the same value of the applied pressure gradient. In two spatial dimensions, the dimensionless equations of motion are given by
\begin{align}
\label{eq:ptt} 
& \frac{\partial {\bm c}}{\partial t} + {\bm v}\cdot\nabla{\bm c} - \left(\nabla {\bm v}\right)^T\cdot{\bm c} - {\bm c}\cdot\left(\nabla {\bm v}\right) = \kappa \nabla^2 {\bm c} \nonumber \\
& \qquad\qquad\qquad\qquad - \frac{{\bm c}-\mathbb{I}}{\Wi}\Bigg[ 1-2\,\epsilon + \epsilon\, \mathrm{Tr} {\bm c}\Bigg],  \\
& \label{eq:ns} 
\frac{\partial {\bm v}}{\partial t} + {\bm v}\cdot\nabla{\bm v}  =
 -\nabla p + \frac{\beta}{Re} \nabla^2{\bm v} \nonumber \\ 
 & \qquad\qquad\qquad\qquad + \frac{(1-\beta)}{Re\,\Wi}\nabla\cdot{\bm c} +
\begin{pmatrix}
           2/Re \\
           0         
 \end{pmatrix}, \\
&\label{eq:incomp} 
\qquad\qquad\qquad\qquad  \nabla\cdot {\bm v} = 0,
\end{align}
where, $p$ is the pressure, $\bm c$ is the polymer conformation tensor, and $\bm v$ is the fluid velocity. 

The parameter space of this model fluid is spanned by four dimensionless numbers \cite{Datta2021preprint}: i) the Weissenberg number, $\Wi=\lambda\,\mathcal{U} / d$, that controls how strongly polymer molecules are stretched in a flow, and, thus, the magnitude of the elastic stresses; here, $\lambda$ is the polymer relaxation time. ii) the Reynolds number, $Re=\rho\,\mathcal{U} d/(\eta_s+\eta_p)$, that gives the relative importance of inertia compared to the viscous stresses; here, $\rho$ is the density of the fluid. Elastic turbulence is characterised by small values of the Reynolds number and throughout this work we set $Re=10^{-2}$. iii) The ratio of the solvent to the total viscosity of the solution, $\beta = \eta_s/(\eta_s + \eta_p)$, that acts as an indirect measure of the polymer concentration in dilute polymer solutions. iv) The strength of shear-thinning, $\epsilon$, that controls how fast elastic stresses grow with $\Wi$. To represent the weakly 
shear-thinning nature of dilute polymer solutions, we set $\epsilon=10^{-3}$.
We note that the choice of $\mathcal{U}$ as the velocity scale implies that the maximum value of the laminar streamwise velocity profile can exceed unity due to shear thinning. The fluid velocity obeys the no-slip boundary condition, $\bm v(x,y=\pm 1,t) = 0$. The boundary conditions for the conformation tensor \cite{liu2013} are obtained by requiring that ${\bm c}(x,\pm 1,t)$ is equal to the values obtained at the boundaries from Eq.\eqref{eq:ptt} with $\kappa=0$.

Simulations are performed with an in-house MPI-parallel code developed within the Dedalus spectral framework \cite{Burns2020} on a rectangular domain $\left[0,L_x\right]\times \left[-1,1\right]$, where $L_x=10$; periodic-boundary conditions are applied in the $x$-direction. All physical fields are represented by a spectral decomposition \cite{Canuto2012} with $256$ Fourier and $1024$ Chebyshev modes. We checked that this resolution is sufficient to achieve numerical convergence. Time-iteration uses a four-stage, third-order implicit-explicit Runge-Kutta method \cite{Ascher1997} with the timestep $dt=5\cdot 10^{-3}$.


{\revision
The conformation diffusion term, $\kappa \nabla^2 {\bm c}$, with $\kappa=5\cdot 10^{-5}$, is added to Eq.\eqref{eq:ptt} of the main text to stabilise the numerics. Its presence is motivated by kinetic theories of dilute polymer solutions, where it stems from a mean-field approximation of Brownian diffusion of individual polymer molecules \cite{Beris1994}, and is required to ensure the formal existence of solutions to polymeric equations of motion \cite{ElKareh1989}. In the past, this term was often employed with a value of $\kappa$ that was orders of magnitude larger than the values predicted by kinetic theories \cite{Sureshkumar1995}, thus calling into question the validity of their results. We stress that the value of $\kappa$ used in this work does not suffer from this drawback. It was selected to be sufficiently large to stabilise the dynamics against short-wave-length numerical instabilities \cite{Owens2002}, yet small enough to be consistent with the kinetic theory values. Indeed, re-casting the kinetic theory prediction in terms of our dimensionless parameters yields $\kappa = D \lambda / d^2 \Wi$, where $D$ is the diffusion coefficient of a polymer molecule in equilibrium. Using $D\sim 1\mu$m$^2$/s, $\lambda\sim 10$s, and $d\sim 100\mu$m, typical for the microfluidic experiments of Pan \emph{et al.} \cite{Pan2013}, and setting $\Wi\sim30$, corresponding to the onset of sub-critical solutions reported below, yields $\kappa\sim 3\cdot 10^{-5}$. While this estimate can yield smaller values of the dimensionless diffusivity for other fluids and wider channels, we have confirmed that our value of $\kappa$ can be further decreased by using higher resolutions and smaller values of $dt$ without altering the results. Similar values of $\kappa$ were previously shown \cite{Page2020} to yield results indistinguishable from the ones obtained with $\kappa=0$.}

All simulations are started from an initial condition comprising the laminar profile and a small localised perturbation \cite{Schumacher2001} to the $xx$-component of the conformation tensor in the following form:
\begin{align}
\Delta \exp{ \left[ -\frac{25}{8} \left\{ \left(\frac{2x}{L_x}-1\right)^2+y^2 \right\} \right] },
\end{align}
where the amplitude $\Delta$ is fixed to be $1\%$ of the maximum value of $c_{xx}$ in the laminar state. {\revision We have verified that other initial conditions, including small random perturbations added to the laminar profile, lead to the same ultimate steady state; however, we found that perturbations in the form presented above are the most efficient in triggering TWS.} Simulations were stopped when the absolute value of the time derivative of the kinetic energy stayed below $10^{-10}$ for more than $10$ time units.

\begin{figure*}
\begin{tabular}{cc}
\begin{tabular}{c}
\includegraphics[width=0.6\textwidth,valign=m]{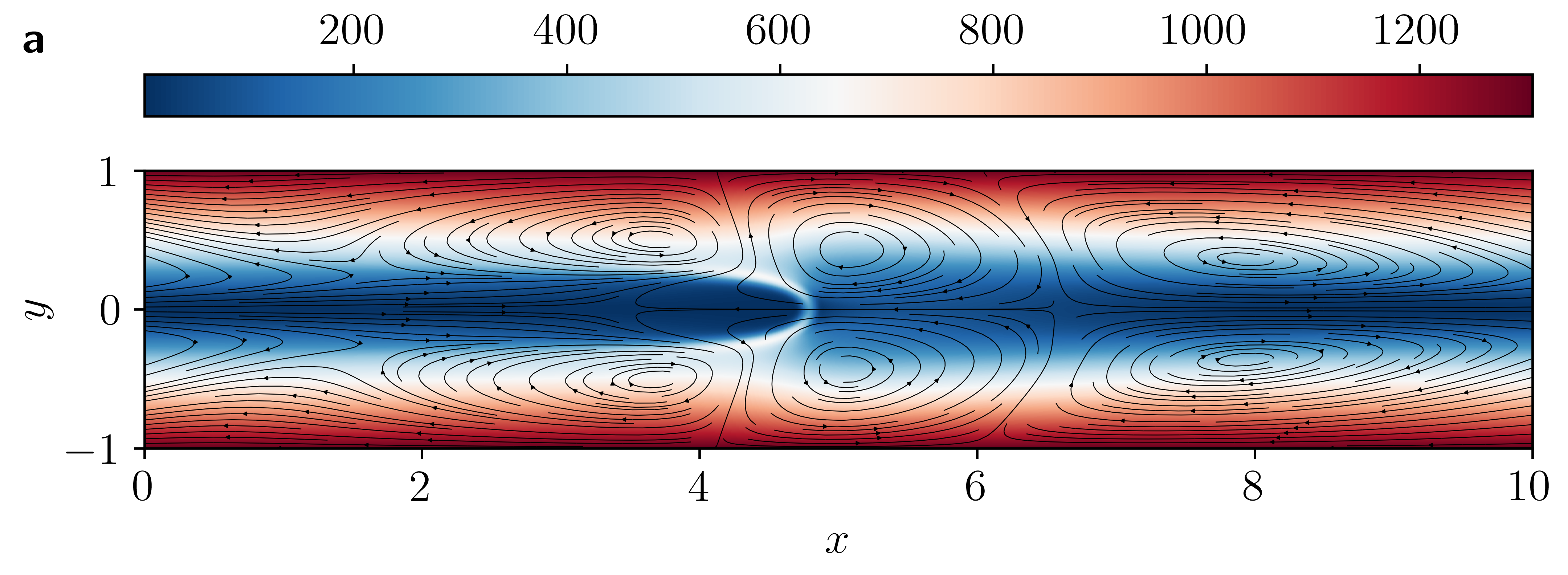} \\ 
\includegraphics[width=0.6\textwidth,valign=m]{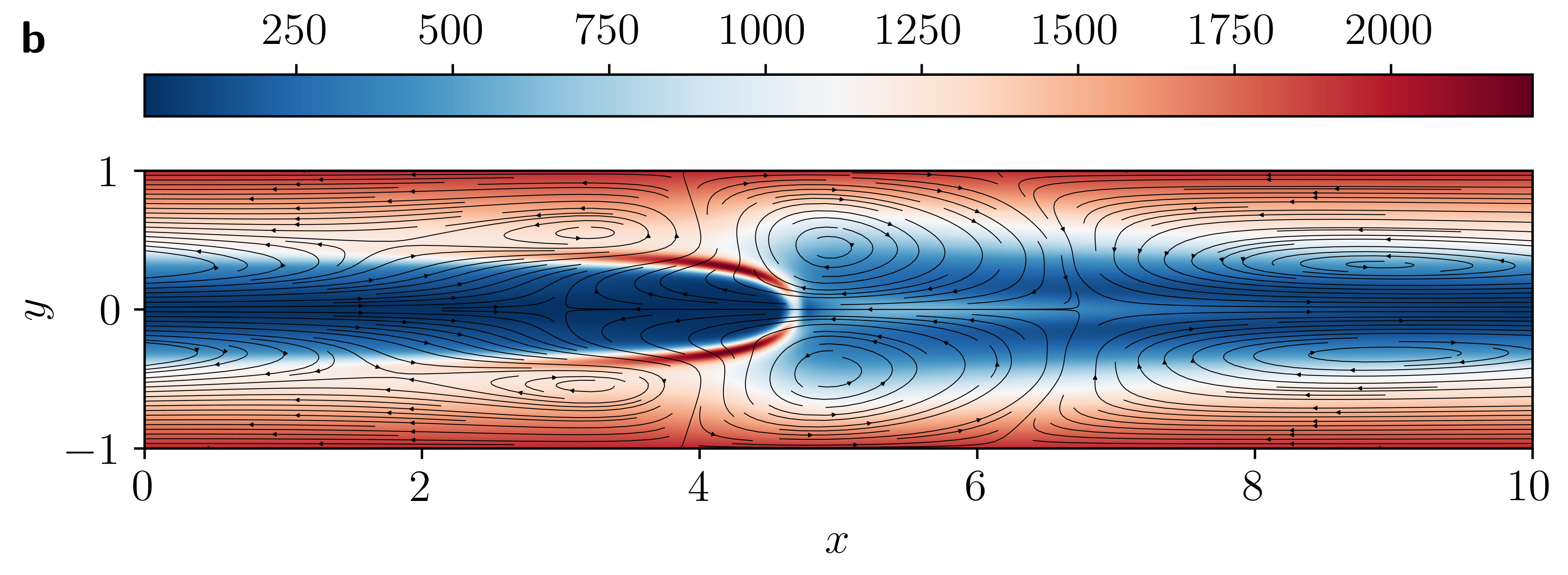} \\ 
\includegraphics[width=0.6\textwidth,valign=m]{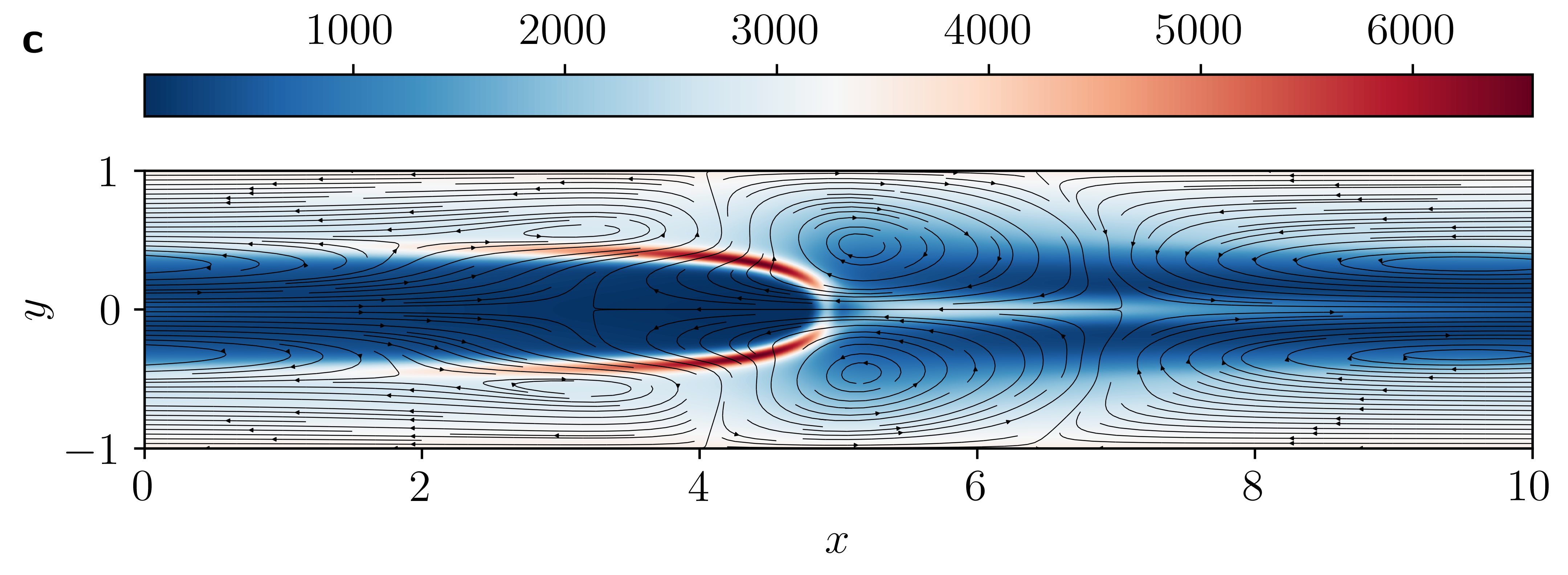}
\end{tabular} 
&
\begin{tabular}{c}
\includegraphics[width=0.3\textwidth,valign=m]{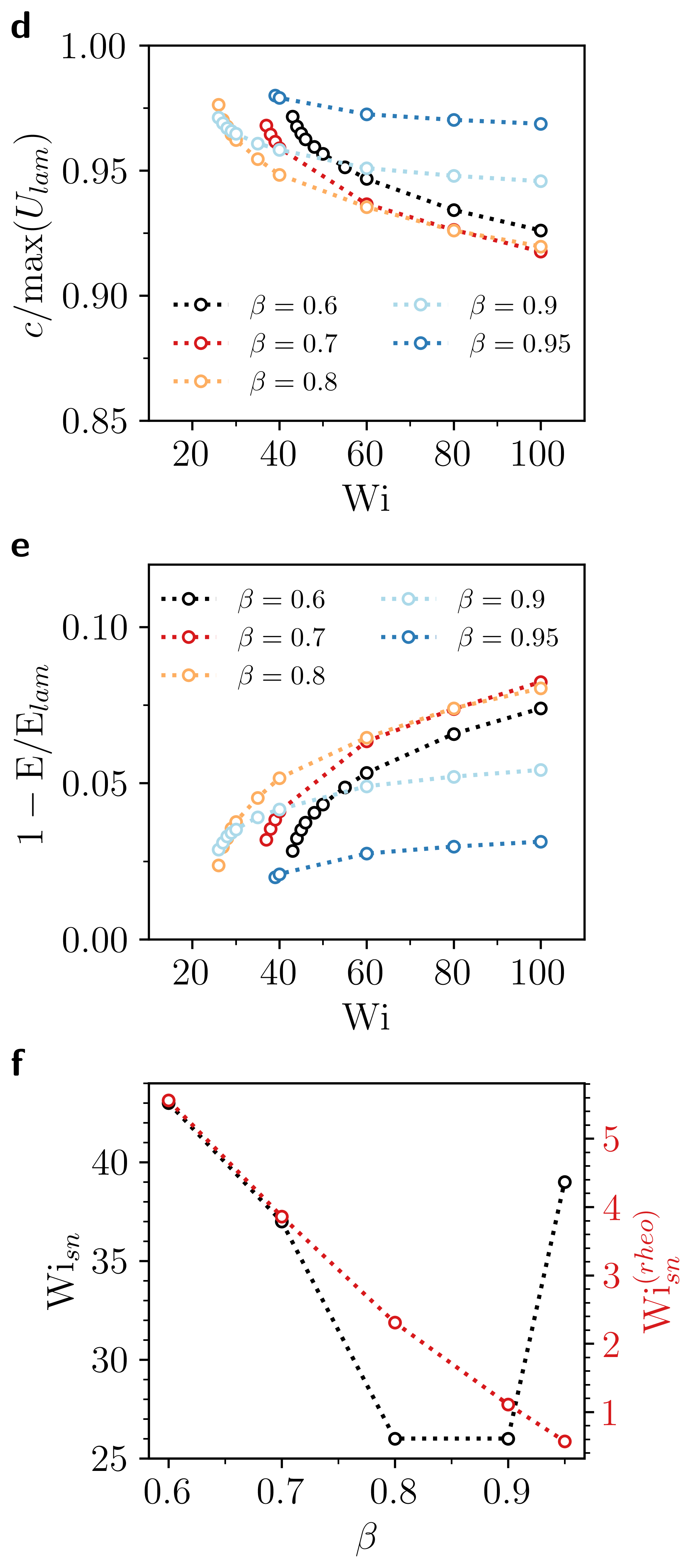}
\end{tabular} 
\end{tabular}
\caption{ Exact travelling-wave solutions. The trace of the conformation tensor (colour) and the flow streamlines (solid lines) for $\beta=0.8$ and $\Wi=26$ ({\bf a}), $\Wi=40$ ({\bf b}), and $\Wi=80$ ({\bf c}). The mean flow is from left to right. The streamlines represent velocity deviation from the mean streamwise profile ${\overline v}_x(y)$.
{\bf d,} The speed of the travelling wave solutions determined by tracking the position of the maximum of $\mathrm{Tr}\,{\bm c}$ as a function of time. {\bf e,}  The relative kinetic energy of the travelling wave solutions tracing the upper branch of the corresponding bifurcation from infinity. {\bf f,} The saddle-node values $\Wi_{sn}$ (black symbols) and their rheological counterparts $\Wi_{sn}^{(rheo)}$ (red symbols) determined as the lowest value of the Weissenberg number at which the solution can be sustained.}
\label{Fig1}
\end{figure*}

Above a critical value of the Weissenberg number, we observe the appearance of purely elastic flow structures that are different from the laminar flow. In Figs.\ref{Fig1}a-c we present examples of such structures for $\beta=0.8$ and $\Wi=26$, $40$, and $80$. The distinctive features of these states are the thin, filament-like arrangements of the polymer stretch, which is proportional to the trace of the conformation tensor $\bf c$. While at lower values of $\Wi$, the maximum polymer extension occurs at the channel walls, as in the laminar profile, at higher $\Wi$, it is localised around the centre of the channel. The arrangements of the stress and velocity profiles in these states move downstream with a constant velocity, i.e. they are steady-states in a co-moving frame and, thus, represent travelling-wave solutions to the equations of motion. In Fig.\ref{Fig1}d, we present their downstream speed $c$, measured by tracking in time the streamwise position of $\mathrm{max(Tr}\,\bm c\mathrm{)}$. Similar to their Newtonian counterparts \cite{Graham2021}, purely elastic TWS move slower than the laminar velocity profile.

Since the laminar flow is linearly stable for the parameters considered here, TWS appear through a sub-critical bifurcation from infinity \cite{Rosenblat1979}. Our direct numerical simulations resolve the upper branches of the bifurcation diagram (Fig.\ref{Fig1}e) that we report in terms of the total kinetic energy of the flow compared to its laminar value at the same $\Wi$ (see \cite{SI} for details). Lower branches, which determine the strength of a finite-amplitude perturbation necessary to destabilise the laminar flow \cite{Morozov2007}, are usually linearly unstable and cannot be determined by direct numerical simulations. The bifurcation diagrams terminate at the saddle-node value of the Weissenberg number, $\Wi_{sn}$, determined as the lowest value of $\Wi$ for which TWS were observed. In Fig.\ref{Fig1}f we map the region of existence of TWS for various values of $\beta$.

\begin{figure*}
\begin{tabular}{ccc}
\includegraphics[width=0.6\textwidth,valign=m]{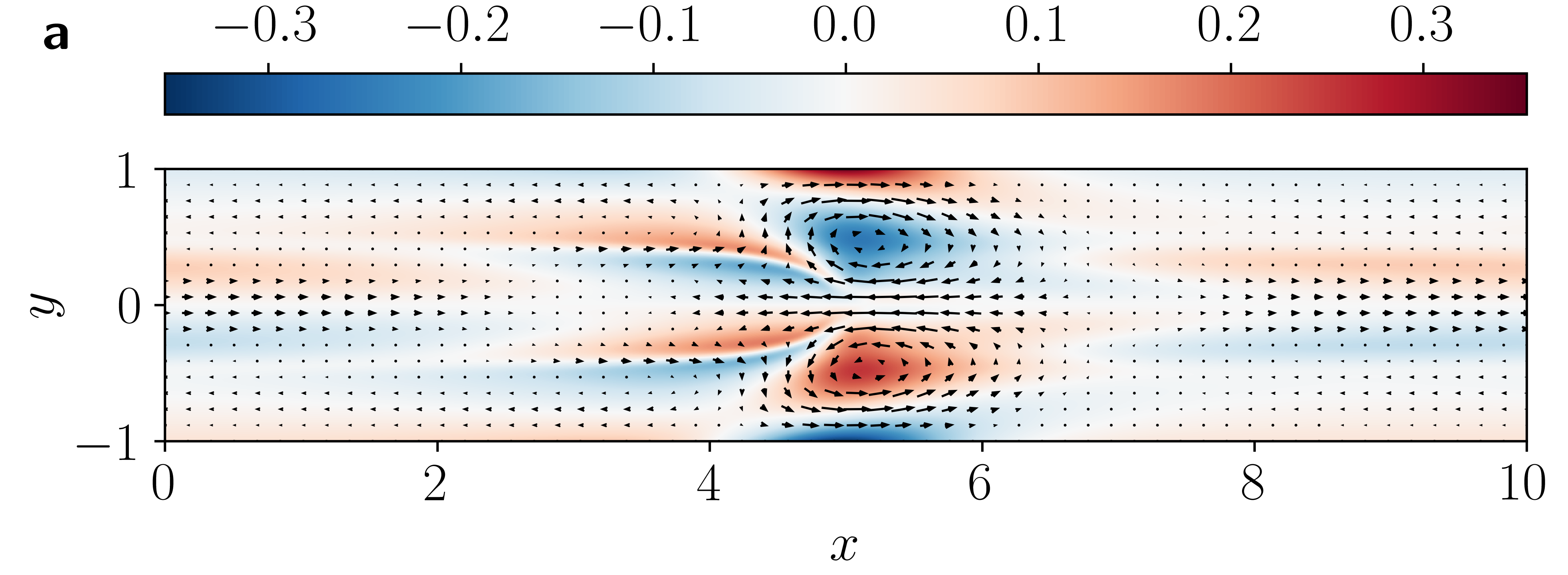} &  &
\includegraphics[width=0.3\textwidth,valign=m]{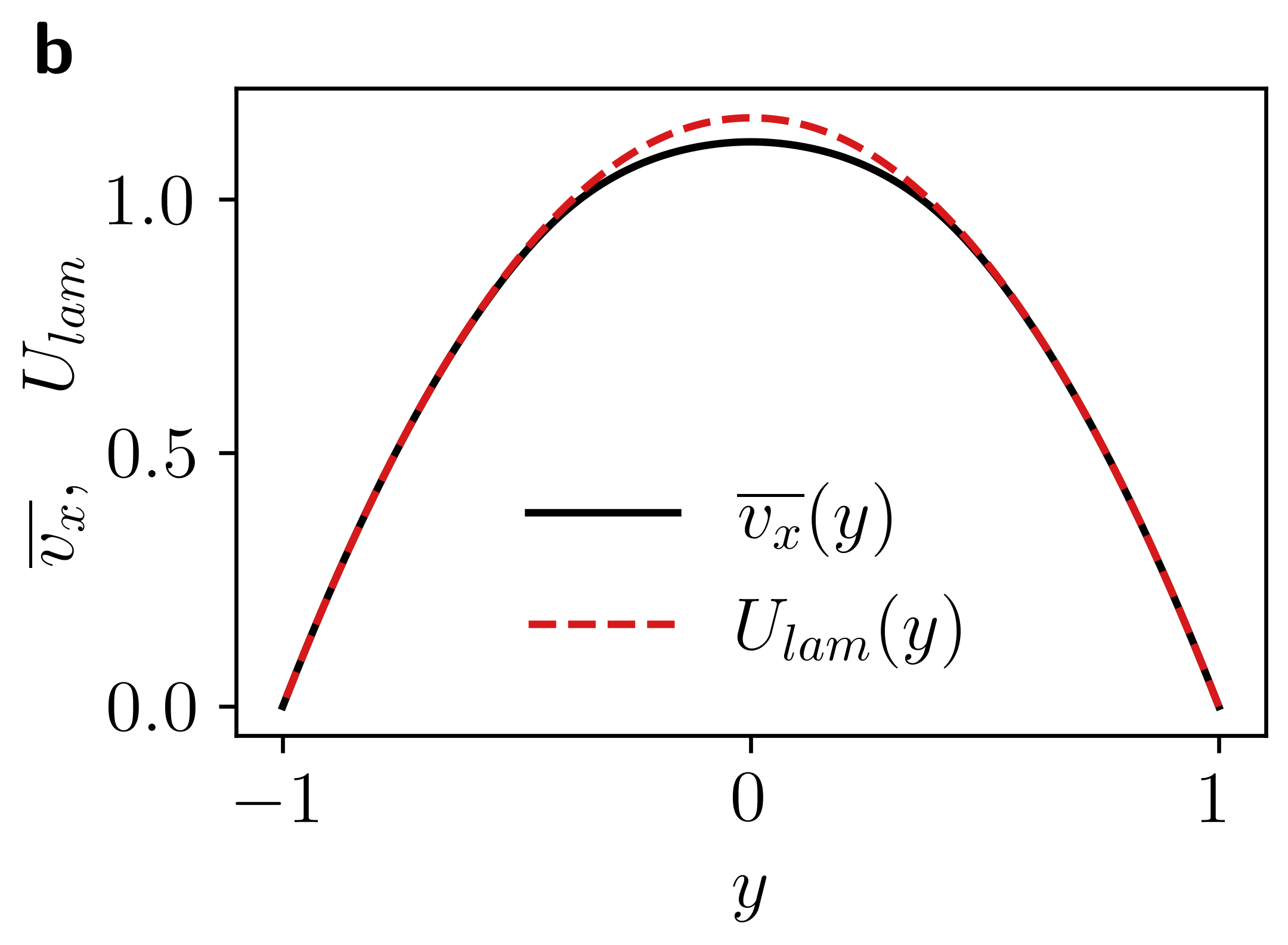}
\end{tabular} 
\caption{ {\revision Characterisation of the TWS velocity field for $\beta=0.8$ and $\Wi=80$. {\bf a,} The velocity deviation from the mean profile $(v_x(x,y)-{\overline v}_x(y),v_y(x,y))$ (vectors), and the out-of-plane component of the vorticity $\partial_x v_y(x,y) - \partial_y (v_x(x,y)-{\overline v}_x(y))$ (colour). 
{\bf b,} Comparison between the laminar and the mean streamwise velocity profiles.}}
\label{Fig2}
\end{figure*}

{\revision To facilitate comparison with experiments, we express $\Wi_{sn}$ in terms of the rheological Weissenberg number $\Wi_{sn}^{(rheo)}$ often used to report the results of experiments dealing with shear-thinning fluids (see, for instance, \cite{McKinley1996,Pan2013,Schaefer2018}). Since the relaxation time and the viscosity of such fluids change with the local shear rate, the nominal value of $\Wi$, which is based on the relaxation time $\lambda$ measured at a single (low) shear rate, is an overestimate of the strength of the elastic stresses. Instead, the rheological Weissenberg number provides an estimate of the value of the Weissenberg number in an Oldroyd-B fluid that would generate elastic stresses of the same magnitude; here, we employ the definition as used by Pan \emph{et al.} \cite{Pan2013} (see also \cite{SI}).  In Fig.\ref{Fig1}f, we report the onset of TWS in terms of the rheological Weissenberg number $\Wi_{sn}^{(rheo)}$. 
In the experiments of Pan \emph{et al.} \cite{Pan2013}, a sudden onset of large velocity fluctuations was observed for $\Wi^{(rheo)}>5.4$; at those flow rates, the total viscosity of the solution was measured to be $\eta_s+\eta_p\sim 0.35$Pa$\cdot$s, while the solvent viscosity $\eta_s \sim 0.2$Pa$\cdot$s, yielding $\beta\sim0.57$ (see Supplemental Material for \cite{Pan2013}). 
{\revisiontwo As can be seen from Figs.1f and 1e, the sub-critical nature of the transition, the value of $\Wi_{sn}$, and the amplitude of the jump of the velocity fluctuations around $\Wi_{sn}$ reported by Pan \emph{et al.} \cite{Pan2013}, are consistent with the onset of TWS found here.}


{\revision
The exact solutions reported in our work resemble the `arrowhead' structures \cite{Dubief2013,Sid2018,Page2020,Dubief2021preprint,buza2021preprint} found in elasto-inertial ($Re\gg1$ and $\Wi\gg1$) pressure-driven channel flows. There, they originate from a linear instability recently discovered for $1-\beta\ll1$, and extend sub-critically to lower values of $Re$ and $\Wi$ \cite{Page2020,buza2021preprint}. This linear instability can be continuously traced to the purely elastic regime \cite{Khalid2021,buza2021preprint}, where, similar to its elasto-inertial analogue, it exists only in a narrow range $1-\beta\ll1$. While such a linear instability is probably not directly relevant for experiments with dilute polymer solutions, as it is difficult to simultaneously achieve $Re\ll1$, $\Wi\gg1$, and $1-\beta\ll1$, we speculate that the non-linear state that originates from it directly connects to TWS discovered here. Importantly, our results demonstrate that no linear instability is required to sustain travelling waves in dilute polymer solutions at experimentally relevant values of $\beta$, and that their origin is purely elastic in nature.}

{\revision Several prominent features of TWS are good potential candidates for experimental detection. The velocity field associated with TWS is dominated by a solitary pair of counter-rotating vortices arranged symmetrically around the centreline of the channel, see Fig.\ref{Fig2}a. Such structures resemble `diwhirls' previously reported in Taylor-Couette flows of dilute polymer solutions \cite{Groisman1997,Kumar2000}. In the absence of other prominent velocity structures, they might be directly observable in particle image velocimetry, although their magnitude is quite low when compared to the mean profile. The latter, defined as ${\overline v}_x(y)=\int_0^{L_x} v_{x}(x,y) dx/L_x$, appears to be the most promising for experimental observation feature of the velocity field, and is the main observable used by Pan {\it et al.} \cite{Pan2013}. However, its deviations from the laminar profile are localised around the centreline and are weak (Fig.\ref{Fig2}b). This is in contrast with Newtonian parallel shear flows, where turbulent production is largely associated with the walls \cite{barkley_2016,Graham2021}. The associated stresses, on the other hand, are about $10\%$ of their laminar values, have a distinctive, filament-like spatial distribution, and should be detectable in birefringence measurements. Finally, a more pronounced, though less specific, signature of TWS is the pressure drop along the channel that can become comparable to its laminar value (see Fig.S2 in \cite{SI}). 

Two-dimensional TWS presented here are steady-states in a co-moving frame and do not lead to time-dependent fluctuations associated with elastic turbulence. To be relevant to the latter, they are expected to lose their stability for sufficiently high $\Wi$. While we have verified numerically that TWS are stable with respect to small two-dimensional perturbations, their stability with respect to three-dimensional ones is yet unknown. There are two possible scenarios. First, the upper branch solutions in Fig.\ref{Fig1}e can be linearly stable up to $\Wi_{3D}$, which would mark the appearance of three-dimensional coherent structures. In this scenario, TWS should be directly observable for $\Wi_{sn}<\Wi<\Wi_{3D}$, and their properties can be tested against our predictions reported in Fig.1. Alternatively, the whole upper branch can be unstable with respect to three-dimensional perturbations. This scenario would mirror the transition to Newtonian turbulence in plane channel flows. There, two-dimensional Tollmien-Schlichting TWS originate from a linear instability at $Re=5772$ and extend sub-critically to lower values of $Re$ \cite{Herbert1976}, while the corresponding upper branch is unstable towards three-dimensional perturbations for any value of $Re$ \cite{Orszag1983}. The close agreement between the onset of velocity fluctuations reported by Pan \emph{et al.} \cite{Pan2013} and the onset of two-dimensional TWS discovered here suggests that $\Wi_{3D}$ is either close or equal to $\Wi_{sn}$. In this case, the flow properties would be determined by three-dimensional structures that are yet to be discovered. Recent studies of elasto-inertial turbulence suggest that such structures are only weakly three-dimensional \cite{Samanta2013,Dubief2013,Sid2018,Shekar2019,Shekar2021,Dubief2021preprint}, and we could expect the two-dimensional flow features discussed above to be observable in three-dimensional channel flows.

Finally, we note that a series of recent experiments from the Steinberg's group \cite{jha2020preprint,Jha2021,shnapp2021preprint,li2022preprint} has suggested a potentially different transition scenario that is rationalised in terms of \emph{elastic Alfv\'{e}n waves}\cite{Varshney2019}. While our results are consistent with our earlier proposal \cite{Morozov2007,Morozov2019}, and with the experiments of the Arratia's group \cite{Pan2013,Qin2017,Qin2019}, further experimental and numerical work is required to understand the transition to elastic turbulence in parallel shear flows. We note that some of the discrepancies between the two sets of observations can be resolved if the aformentioned elastic waves are interpreted as TWS discussed here.}


\section*{Acknowledgements}

This work was partially funded by  EPSRC (grant number EP/I004262/1). We would like to thank EPSRC for the computational time made available on the ARCHER2 UK National Supercomputing Service (https://www.archer2.ac.uk) via the UK Turbulence Consortium (EP/R029326/1). We thank Paulo Arratia, Martin Lellep, Moritz Linkmann, Davide Marenduzzo, Jacob Page, and Rob Poole for helpful discussions. We gratefully acknowledge the support we received from the Dedalus team (https://dedalus-project.org), and Keaton Burns in particular. 

\bibliographystyle{apsrev4-2}
\bibliography{nature}

\end{document}


\title{Coherent structures in plane channel flow of dilute polymer solutions \\[2mm] Supplemental Material}

\author{Alexander Morozov}

\maketitle

\subsection*{Laminar profile and kinetic energy}

The laminar state is defined by ${\bm v}_{lam} = \left(U_{lam}(y),0,0\right)$ and 
\begin{align}
{\bm c}_{lam} = \begin{pmatrix}
a_{xx}(y) & a_{xy}(y) \\
a_{xy}(y) & 1
\end{pmatrix},
\end{align}
where the velocity profile $U_{lam}(y)$ and the components of the conformation tensor satisfy the one-dimensional version of Eqs.(1)-(3):
\begin{align}
&\label{eq:lamaxx}  
\frac{a_{xx}-1}{\Wi}\Big[ \epsilon\left(a_{xx}-1\right)+1\Big] - \kappa a_{xx}'' = 2 a_{xy}U_{lam}', \\
&\label{eq:lamaxy} 
\qquad \frac{a_{xy}}{\Wi}\Big[ \epsilon\left(a_{xx}-1\right)+1\Big] - \kappa a_{xy}'' = U_{lam}', \\
&\label{eq:lamns} 
\qquad\qquad \beta U_{lam}'' + \frac{1-\beta}{\Wi} a_{xy}' + 2 = 0.
\end{align}
The velocity profile satisfies $U_{lam}(\pm 1) = 0$, while $a_{xx}(\pm 1)$ and $a_{xy}(\pm 1)$ are set to their corresponding values obtained by solving Eqs.\eqref{eq:lamaxx}-\eqref{eq:lamns} with $\kappa=0$. \\

The ratio of the instantaneous and laminar kinetic energies is defined as
\begin{align}
\frac{\textrm{E}}{\textrm{E}_{lam}} = \frac{ \frac{1}{L_x}\int_{0}^{L_x} dx \int_{-1}^{1}dy \left( v_x^2 + v_y^2 \right) }{ \int_{-1}^{1}dy\, U_{lam}^2 }.
\end{align}

\subsection*{Linear stability analysis}

Stability of the laminar flow is determined by studying time evolution of infinitesimal disturbances. To this end, we introduce a perturbation to the laminar profile in the following form
\begin{align}
& \left( {\bm c}, {\bm v}, p\right)(x,y,t) = \left( {\bm c}_{lam}, {\bm v}_{lam}, 0\right)(y) \nonumber \\
& \qquad\qquad\qquad\qquad +  e^{i k x} e^{\sigma t}\left( \delta{\bm c}, \delta{\bm v}, \delta p\right)(y),
\end{align}
where $k$ sets the perturbation's periodicity in the $x$-direction, and $\sigma$ is a yet to be determined temporal eigenvalue. To first order, the perturbation obeys the linearised Eqs.(1)-(3) that we solve numerically using a spectral method based on Chebyshev polynomials \cite{Canuto2012}. 

\begin{figure*}[!ht]
\includegraphics[width=\textwidth,valign=m]{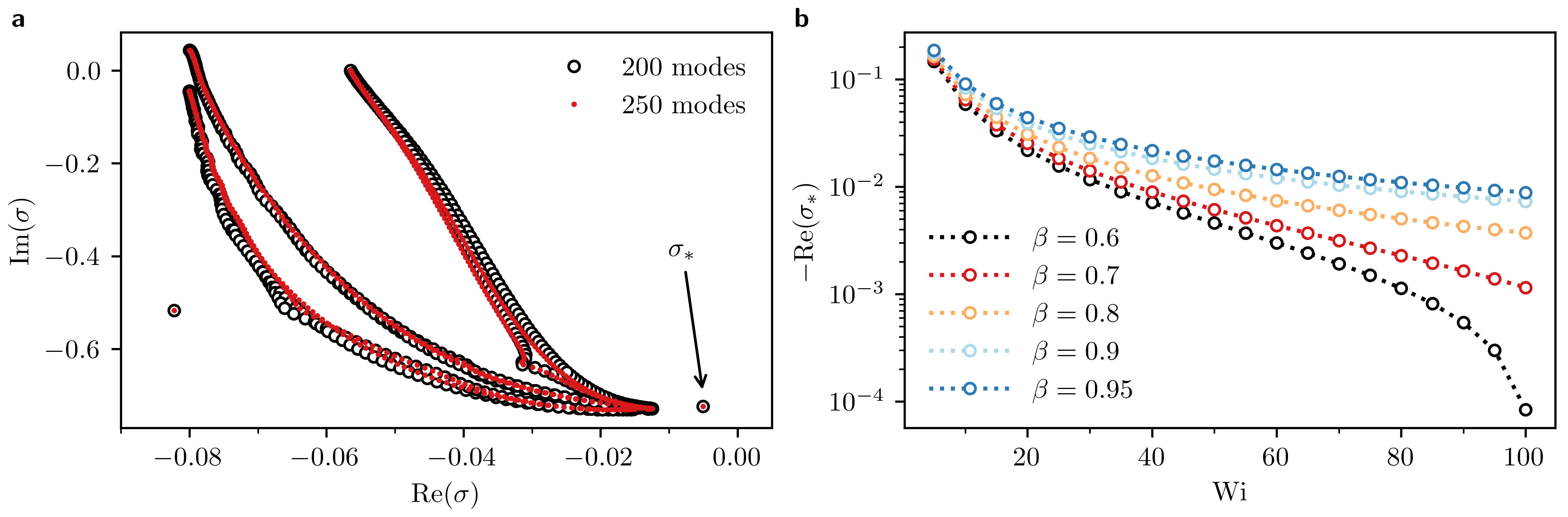}
\caption{ The results of linear stability analysis for $k=2\pi/L_x$. {\bf a,} The eigenvalue spectrum for $\beta=0.8$ and $\Wi=80$ at two Chebyshev resolutions, showing numerical convergence. The leading eigenvalue is denoted by $\sigma_*$. {\bf b,} The real part of the leading eigenvalue as a function of $\Wi$ for various values of $\beta$ calculated with $250$ Chebyshev modes. No linear instability is found.}
\label{FigSLinearStability}
\end{figure*}

For all values of $\Wi$ and $\beta$, we find that the real part of $\sigma$ is always negative, i.e. the laminar flow is linearly stable (Fig.\ref{FigSLinearStability}), confirming the bifurcation-from-infinity scenario for the appearance of the travelling-wave solutions reported in the main text. These results are in line with the previous work on linear stability of the Oldroyd-B \cite{Zhang2013,Khalid2021} and FENE-P \cite{Zhang2013,buza2021preprint} models; the latter is particularly relevant to this work due to the intrinsic relationship between the simplified PTT and FENE-P models \cite{Poole2019}.

\subsection*{Rheological Weissenberg number}

To assess the relative strength of the polymeric stresses at a particular shear rate $\dot\gamma$, we introduce the rheological Weissenberg number $\Wi^{(rheo)}$. As mentioned in the main text, it provides an estimate of the value of the Weissenberg number in an Oldroyd-B fluid that would generate elastic stresses of the same magnitude and serves as a phenomenological way of factoring the shear-rate dependence of the fluid properties out of the definition of the Weissenberg number. Here, we employ the definition used by Pan \emph{et al.} \cite{Pan2013}:
\begin{align}
\Wi^{(rheo)} = \frac{N_1(\dot\gamma)}{2 \left[ \tau_{xy}(\dot\gamma)+\eta_s \dot\gamma\right]},
\end{align}
where $N_1$ and $\tau_{xy}$ are the polymeric contributions to the first normal-stress difference and shear stress in simple shear flow, respectively. When used for an Oldroyd B fluid, this definition gives $\Wi^{(rheo)}=Wi$. When instead adapted for the linear Phan-Thien-Tanner model, it yields
\begin{align}
\Wi^{(rheo)} = \frac{c_{xy}^2}{c_{xy}+\frac{\beta }{1-\beta}\Wi},
\end{align}
where $c_{xy}$ in simple shear is given by
\begin{align}
c_{xy} \left[ 1+ 2\, \epsilon \, c_{xy}^2\right] = \Wi.
\end{align}
For $\Wi\ll (2\epsilon)^{-1/2}$, $\Wi^{(rheo)}\sim \Wi$, while for $\Wi\gg (2\epsilon)^{-1/2}$, $\Wi^{(rheo)}\sim \Wi^{-1/3}$, indicating the shear-thinning induced weakening of the elastic stresses at large $\Wi$.

\section*{Supplementary Figures}

\begin{figure*}[!ht]
\includegraphics[width=0.6\textwidth,valign=m]{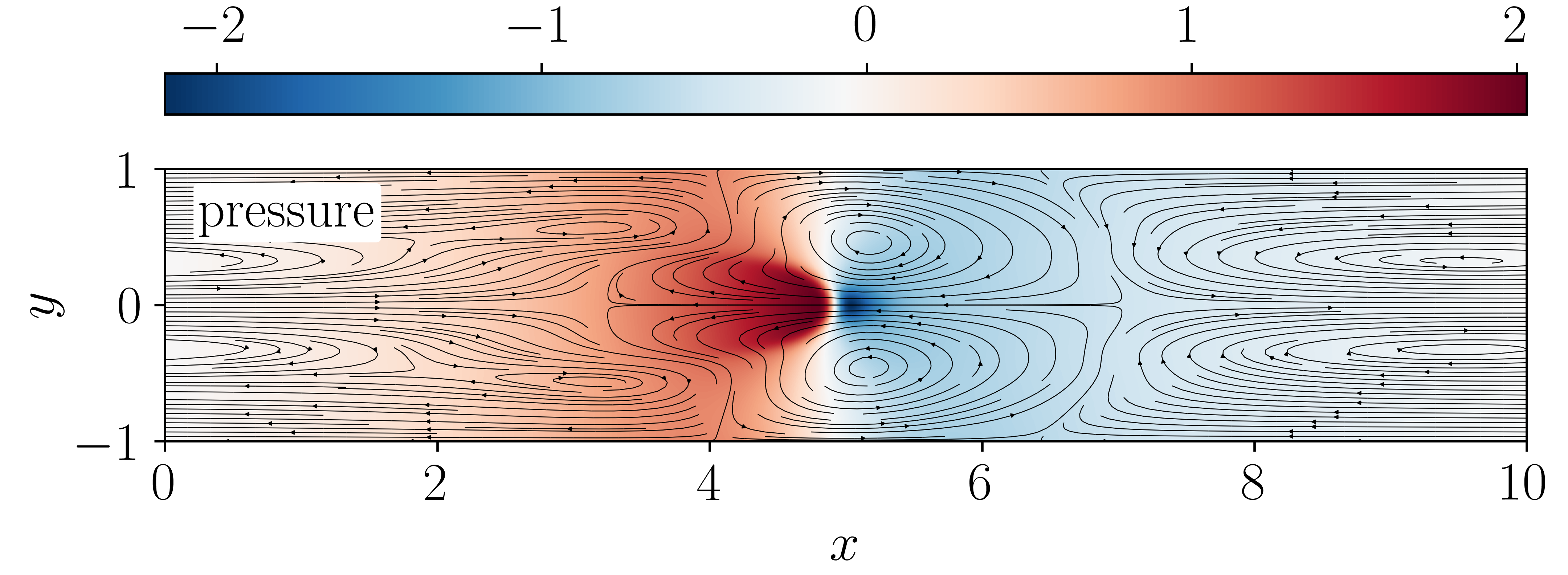}
\caption{ Normalised pressure profile $p\,Re$ of the travelling wave solution for $\beta=0.8$ and $\Wi=80$. In these units, the laminar pressure gradient along the channel is $2$.}
\label{FigS2}
\end{figure*}

\newpage

\bibliographystyle{apsrev4-2}
\bibliography{nature}